%% file: main.tex
\newcommand{\CLASSINPUTtoptextmargin}{0.78in}
\documentclass[conference]{IEEEtran}
\IEEEoverridecommandlockouts

\usepackage{amsmath,amssymb,amsfonts}
\usepackage{algorithmic}
\usepackage{graphicx}
\usepackage{textcomp}
\usepackage{xcolor}
\usepackage{bbm, dsfont}

\def\BibTeX{{\rm B\kern-.05em{\sc i\kern-.025em b}\kern-.08em
    T\kern-.1667em\lower.7ex\hbox{E}\kern-.125emX}}

\input{user_macros_packages.tex}

\begin{document}

\title{An Efficient Synchronization Scheme for Distributed Bandits

\thanks{This paper is supported by the CHIST-ERA Wireless AI 2022 call MLDR project (ANR-23-CHR4-0005), partially funded by AEI and NCN under projects PCI2023-145958-2 and 2023/05/Y/ST7/00004, respectively.}
\thanks{\textcopyright~2026 IEEE.  Personal use of this material is permitted.  Permission from IEEE must be obtained for all other uses, in any current or future media, including reprinting/republishing this material for advertising or promotional purposes, creating new collective works, for resale or redistribution to servers or lists, or reuse of any copyrighted component of this work in other works.}
}

\author{\IEEEauthorblockN{Raymond Zhang, Henrique Miyamoto, Richard Combes, Sheng Yang}
\IEEEauthorblockA{\textit{Laboratoire des Signaux et Systèmes (L2S)} \\
\textit{Université Paris-Saclay, CNRS, CentraleSupélec}\\
Gif-sur-Yvette, France\\
\{raymond.zhang, henrique.miyamoto, richard.combes, sheng.yang\}@centralesupelec.fr}
}

\maketitle

\begin{abstract}
  We study the problem of synchronizing two players in a \emph{distributed multi-armed bandit} setting. The probability of \emph{synchronization error} is analyzed by recasting synchronization as a \emph{communication problem}. Using a well-designed random coding ensemble, we derive error exponents for the maximum-likelihood decoder. We then introduce a \emph{universal decoder} that achieves the same error exponents as the optimal decoder. Finally, applying these results to distributed multi-armed bandits, we show an upper bound\footnote{The minimax lower bound (Theorem 7) in our conference paper 10.1109/ISIT62367.2026.11654092 is incorrect. The averaging over the prior step does not hold for the event that we consider. Some algorithms have provably lower initialization regret on the class of problems considered.} on the regret of our synchronization scheme. 
\end{abstract}

\begin{IEEEkeywords}
Multi-Armed Multiplayer Bandits, Random Codes, Universal Decoding 
\end{IEEEkeywords}

\input{1_introduction.tex}

\input{2_bandit_channel.tex}
\input{3_random_codes.tex}

\input{4_universal_decoder.tex}

\input{5_application_to_bandit.tex}

\input{6_numerical_results.tex}

\bibliographystyle{IEEEtran}
\bibliography{references}

\end{document}

%% file: user_macros_packages.tex
\usepackage{amsmath}
\usepackage{amsthm}
\usepackage{bm}
\usepackage{caption}
\usepackage{subcaption}
\usepackage{tcolorbox}
\usepackage{rotating}
\usepackage{scalerel}
\usepackage{float}
\usepackage{multirow}
\usepackage{diagbox}
\usepackage{enumitem}
\usepackage{subfiles}

\usepackage[linesnumbered,ruled,vlined,boxed]{algorithm2e}

\SetKwInput{KwInit}{Init}

\renewcommand{\epsilon}{\varepsilon}

\renewcommand{\hat}{\widehat}

\DeclareMathOperator{\argmax}{argmax}

\theoremstyle{plain}
\newtheorem{question}{\protect\questionname}
\providecommand{\questionname}{Question}

\theoremstyle{plain}
\newtheorem{questionfr}{\protect\questionname}

\theoremstyle{plain}
\newtheorem{theorem}{\protect\theoremname}
\theoremstyle{plain}
\newtheorem{cor}[theorem]{\protect\corname}
\theoremstyle{plain}

\theoremstyle{plain}

\theoremstyle{plain}

\theoremstyle{plain}

\providecommand{\algorithmname}{Algorithm}

\providecommand{\lemmaname}{Lemma}
\providecommand{\theoremname}{Theorem}
\providecommand{\corname}{Corollary}
\providecommand{\propositionname}{Proposition}
\providecommand{\remarkname}{Remark}

\newcommand{\Biid}{\text{B-iid}}
\newcommand{\Bcc}{\text{B-cc}}

\definecolor{ududff}{rgb}{0.30196078431372547,0.30196078431372547,1}
\definecolor{xdxdff}{rgb}{0.49019607843137253,0.49019607843137253,1}
\definecolor{uuuuuu}{rgb}{0.26666666666666666,0.26666666666666666,0.26666666666666666}

\newcommand{\hq}[1]{#1}
\makeatletter
\DeclareRobustCommand*\cal{\@fontswitch\relax\mathcal}
\makeatother
\usepackage{comment}

\usepackage{mathtools}

\newcommand{\xv}{\pmb{x}}
\newcommand{\yv}{\pmb{y}}

\newcommand{\Xv}{\pmb{X}}
\newcommand{\Yv}{\pmb{Y}}

\newcommand{\rw}{\mu}

\renewcommand{\leq}{\leqslant}
\renewcommand{\geq}{\geqslant}

\newcommand{\given}{\mid}
\newcommand{\mgiven}{\;\middle\vert\;}

\newcommand{\1}{\mathbbm{1}}

\newcommand{\Acal}{\mathcal{A}}

\newcommand{\Xcal}{\mathcal{X}}
\newcommand{\Ycal}{\mathcal{Y}}
\newcommand{\Bv}{\pmb{B}}
\newcommand{\Qv}{\pmb{Q}}
\newcommand{\ML}{\text{ML}}

\renewcommand{\bar}{\overline}
\renewcommand{\tilde}{\widetilde}

%% file: 1_introduction.tex
\section{Introduction}
\label{sec:introduction}

The main motivation for this work is multi-player multi-armed bandits, a repeated game between $J$ players sharing $K$ arms during $T$ rounds (see the survey~\cite{boursier_survey_2024} for a complete overview). 
At each round, based on prior observations, each player $j \in [J]$ chooses an arm $x_j \in [K]$ and observes a Bernoulli reward with mean $\rw_{x_j} C_j(x) $, with $C_j(x) \coloneqq \prod_{j' \ne j} \1 \left\{ x_{j'} \ne x_{j} \right\}$ the collision indicator, and $x \coloneqq (x_{1},...,x_J)$ the vector of chosen arms, and where the vector $\rw \coloneqq \left( \mu_{1}, \dots, \mu_{K} \right)$ is initially unknown to players.
It is noted that if there exists a player $j' \ne j$ such that $x_j = x_{j'}$ we say that $j$ and $j'$ \emph{collide}; in turn, $C_j(x) = 0$ and the reward of player $j$ is $0$. 
The goal is, without any coordination between players or prior information, and solely by observing rewards, for players to learn the decision $x$ maximizing the total expected reward $\sum_{j=1}^{J} \rw_{x_j} C_j(x)$.
Assume that $J \le K$ and denote by $\rw_{(1)} \ge \dots \ge \rw_{(K)}$ the entries of $\rw$ sorted in decreasing order.
The optimal decision is that each player plays a distinct arm (to avoid collisions) in the set of the $J$ best arms, which yields a total reward of $\sum_{j=1}^J \rw_{(j)}$. 

More precisely, the goal of this problem is to minimize the \emph{regret}, defined as the expected difference between the cumulative sum of rewards obtained by a centralized oracle only selecting the optimal decision, and the cumulative sum of rewards obtained by the players, that is,
\begin{equation}
    \mathcal{R}(T) \coloneqq T \sum_{j=1}^J \rw_{(j)} - \mathbb{E}\left[ \sum_{t=1}^T \sum_{j=1}^J \mu_{X_{t,j}} C_j(X_t)\right],
\end{equation}
where $X_{t,j}$ is the arm chosen by player $j$ at time $t$, and $X_t \coloneqq (X_{t,1}, \ldots, X_{t,J})$ the corresponding vector. 

This problem is known to be non-trivial, and all state-of-the-art algorithms with strong theoretical performance guarantees proposed in~\cite{boursier_sic-mmab_2019,shi_decentralized_2020,wang_optimal_2020,huang22a} use elaborate schemes that leverage collisions to coordinate the players and exchange information between them.
Those schemes allow, for instance, to: elect a leader amongst players, assign arms to players, estimate the number of players, and exchange estimates of~$\rw$ between players.
The simplest of those schemes is called ``Musical Chairs'' (MC)
\cite[Algorithm~1]{rosenski_multi-player_nodate}, where each player samples arms uniformly at random until they see a non-zero reward, and then keep playing that arm for a certain number of rounds. 
Given enough time, this assigns a unique arm to any given player with high probability, which serves as a unique identifier.  
Upon termination, players can then exchange their respective identifiers using another procedure called ``Sequential Hopping'' (SH), see \cite[Algorithm~6]{boursier_sic-mmab_2019}.
The combination of MC followed by SH allows players to synchronize and avoid collisions. Once synchronized, players alternate between exploring the various arms in a distributed fashion, and communicating in order to share those estimates between them.

The regret of state-of-the-art algorithms can be decomposed as \cite[p.~12]{boursier_sic-mmab_2019}:
\begin{equation}
    \mathcal{R}(T) = \mathcal{R}_{\text{init}}(T) + \mathcal{R}_{\text{explor}}(T) +  \mathcal{R}_{\text{comm}}(T).
\end{equation}
The term $\mathcal{R}_{\text{init}}(T)$ is the regret due to the initialization phase during which players get synchronized; $\mathcal{R}_{\text{explor}}(T)$ is the regret incurred when players explore the various arms in a distributed fashion to identify the best arms and estimate $\rw$; and $\mathcal{R}_{\text{comm}}(T)$ is the regret caused by communication phases where players exchange their estimates of~$\rw$.

In this work, we focus on the initialization phase for two players ($J = 2$), and propose efficient schemes to synchronize the players with low regret. 
We propose a novel synchronization scheme that casts the synchronization problem in multi-player multi-armed bandits as a problem of universal decoding~\cite{lapidoth1998b} over the so-called bandit channel~\cite{zhang2025}, and whose performance is analyzed using random coding arguments and error-exponent techniques.
 The regret bound of our algorithm compared to the state of the art (up to universal multiplicative constants) is presented in Table~\ref{tab:regret_init_existing}, along with the assumptions necessary for the bounds to hold. 

\begin{table}[t]
    \caption{Regret of the initialization phase for $J=2$.}
	\begin{center}
		\renewcommand{\arraystretch}{1.2}
		\begin{tabular}{|c|c|c|}
			\hline 
			Reference & Knowledge & $\lim_{T\to+\infty} \frac{\mathcal{R}_{\text{init}}(T)}{\log(T)}$ \\ \hline \hline
			\cite[Lem.~11]{boursier_sic-mmab_2019}, \cite[Lem.~1]{shi_decentralized_2020} & $\mu_{(K)}$ & $K/\mu_{(K)}$\\
			\hline 
			\cite[Lem.~1]{huang22a} & $\emptyset$ & $K^2$ \\
			\hline 
			This work & $\mu_{(1)}$ or $E(\mu)$ & $K$ \\
			\hline 
		\end{tabular}%
    \label{tab:regret_init_existing}
	\end{center}
\end{table}

\paragraph*{Our Contributions}
  1)~We study a random-coding scheme for the bandit channel that does not require knowledge of the channel parameters at the transmitter, and analyze its performance under maximum-likelihood (ML) decoding when the receiver knows the channel parameters. 
  2)~We then consider the case where the receiver does not know the channel parameters, and analyze the decoding error probability of a universal decoder. 
  3)~We apply the results of 1) and 2) to the synchronization problem in multi-player multi-armed bandits, and show that the resulting synchronization scheme achieves the best known regret upper bound, strictly improving upon all previously known schemes, to the best of our knowledge. 
 4)~Finally, we validate our theoretical findings through numerical experiments. 
 Due to space limitations, the detailed proofs of our results are omitted and will be provided in a longer version of this paper. 
\

\paragraph*{Notation} 

Let $\mathcal{P}(\mathcal{X})$ denote the set of probability distributions over a finite alphabet $\mathcal{X}$, $\mathcal{P}_n(\mathcal{X})$ the set of $n$-length types over~$\mathcal{X}$, and $\mathcal{T}_n(Q_X)$ the type class associated with a type $Q_X \in \mathcal{P}_n(\mathcal{X})$. For a sequence $\xv$ and a symbol $x \in \mathcal{X}$, let $a_{\xv}(x)$ denote the number of occurrences of $x$ in~$\xv$. We refer to~\cite[Chapter~2]{csiszar2011} for details on the method of types.
We denote by $\mathbbm{1}\{P\}$ the indicator function that takes the value $1$ if the proposition $P$ is true and $0$ otherwise, and define $\mathbbm{1}_A(x) \coloneqq \mathbbm{1}\{x \in A\}$. We adopt the convention that $0^0 = 1$. The logarithm $\log(\cdot)$ is taken to base~$e$.  We denote $[n] \coloneqq \{1,\dots,n\}$.
For the case of $J=2$ players, given a player $j$, we denote the opposing player by $\sigma(j)$, with $\sigma(1) = 2$ and $\sigma(2) = 1$.

%% file: 2_bandit_channel.tex
\section{Bandit Channel and Synchronization Scheme}
\label{sec:bandit_channel_model}

Consider $J=2$ players, playing a multi-player multi-armed bandit with unknown parameters $\rw$, and define the synchronization problem as follows. 
Initially, both players receive some prior information about $\mu$, and the time horizon~$T$. 
They then choose a synchronization length $n \le T$. 
At times $t=1,...,n$, player $j$ selects arm $X_{t,j}$ and observes a reward $Y_{t,j}$. 
At time $t = n+1$, player $j$ selects their identifier $Z_{j} = f(X^n_j,Y^n_j) \in [J]$. 
The players are said to achieve synchronization if all identifiers are distinct, so that $Z_{1} \ne Z_2$. Upon successful synchronization, the identifiers allow the players to avoid collisions during the remaining rounds $t = n+1, \ldots, T$.

We aim to minimize the synchronization regret, defined as 
\begin{align*}
	\mathcal{R}_{\text{init}}(\mu,T) &=
 n \sum_{j=1}^J \rw_{(j)} - \mathbb{E}\left[ \sum_{t=1}^T \sum_{j=1}^J \mu_{X_{t,j}} C_j(X_t)\right] \\ &+ (T-n)\mathbb{P}\left\{ Z_1 = Z_2 \right\} \sum_{j=1}^J \rw_{(j)},
\end{align*}
where the \hq{last} term penalizes failures to achieve synchronization.

We cast synchronization as a communication problem over the so-called bandit channel, as defined in~\cite{zhang2025}. 
The bandit channel has two inputs and two outputs~(one for each player) with input alphabet $\Xcal \coloneqq [K]$, and output alphabet $\Ycal \coloneqq \{0,1\}$.
The players share a common codebook $\mathcal{C} \coloneqq \left( \xv^{(1)}, \dots, \xv^{(M)} \right) \subseteq \Xcal^{n}$ of blocklength~$n$ and rate $R \coloneqq (\log M)/n$ consisting of $M$ sequences of arm selections. 
Player~$j \in [J]$ chooses a message $M_j \in [M]$, and encodes it by playing the sequence of arms specified by the corresponding codeword $\Xv_j \coloneqq \xv^{(M_j)} \coloneqq \big( x_1^{(M_j)}, \dots, x_n^{(M_j)} \big)$. Player~$j$ then observes the sequence of outputs $\Yv_j \in \Ycal^n$, corresponding to the realized rewards. 
In the absence of output feedback, the memoryless stationary channel has the following transition probability
	\begin{align*} 
		&P_{\Yv_{1},\Yv_{2} \given \Xv_{1},\Xv_{2}}\left(\yv_{1},\yv_{2} \given \xv_{1},\xv_{2}\right) 
		\\
		& \quad = \prod_{t\in[n]}
		P_{Y \given  X_1,X_2} \left(y_{t,1} \vert x_{t,1}, x_{t,2}\right)
		P_{Y \given X_1,X_2} \left(y_{t,2} \vert x_{t,2}, x_{t,1}  \right),
	\end{align*}
	where the single-letter channel law is given by
	\begin{equation} \label{eq:channel-law-single-letter}
		P_{Y\given X_1,X_2}(y|x_1,x_2) \coloneqq
			\begin{cases} 
			 \rw_{x_1} \1\{x_1\ne x_2\} & \text{ if } y = 1  \, ,\\
			  1 - \rw_{x_1} \1\{x_1\ne x_2\} & \text{ if } y = 0  \, ,  \\
			\end{cases} 
	\end{equation}
for $j \in [J]$.
Player $j$ decodes the message of the other player $\sigma(j)$ by computing an estimate $\hat{M}_{\sigma(j)}$ as a function of its input sequence $\Xv_{j}$ and output sequence $\Yv_j$, using the decoding rule
		\begin{equation} \label{eq:decoding-rule}
			\hat{M}_{\sigma(j)} = \argmax_{m\in[M]\setminus m_j} u\left( \Yv_j, \Xv_j; \xv^{(m)} \right),
		\end{equation}
for some metric $u \colon \Ycal^n \times \Xcal^n \times \Xcal^n \to \mathbb{R}$, with ties broken arbitrarily. 
Decoding is said to be successful if $\hat{M}_{\sigma(j)} = {M}_{\sigma(j)}$ for all $j \in [J]$.
If the players know $\rw$, the optimal rule is maximum likelihood, with metric
		\begin{equation} \label{eq:ML-decoding-metric}
			u_{\text{ML}}(\yv,\xv_1;\xv_2)
			= \prod_{t\in[n]} P_{Y_j \given X_1, X_2}\left( y_{t,j} \given x_{t,1}, x_{t,2} \right) \, .
		\end{equation}
		When the players do not know $\rw$, we will propose \emph{universal} decoding rules in which $u$ is independent of $\rw$, which is one of the main contributions of this work. 
	
To synchronize the players using the bandit channel, let each message $M_j$ be chosen independently and uniformly at random from $[M]$. Both players transmit their respective messages, perform decoding, and then select their identifiers by ranking the messages such that $Z_j = 1$ if $M_j \le \hat{M}_{\sigma(j)}$, and $Z_j = 2$ otherwise. 
Synchronization succeeds if the messages $M_1$ and $M_2$ are distinct, and decoding succeeds if $\hat{M}_{\sigma(j)} = {M}_{\sigma(j)}$ for all $j \in [J]$. Consequently, the probability of synchronization error can be bounded as
\begin{align}
	& \hspace{-1em}\mathbb{P} \left\{ Z_1 = Z_2 \given \mathcal{C} \right\} \nonumber\\
	& \leq \mathbb{P}\left\{ \left(\cup_{j\in\{1,2\}} \{M_j \ne \hat{M}_j\}\right) \cup \{M_1 = M_2 \} \given \mathcal{C} \right\} \nonumber\\
	&\le 2 \underbrace{\mathbb{P}\left\{ M_1 \ne \hat{M}_1 \mgiven \mathcal{C}, M_1 \neq M_2 \right\}}_{\eqqcolon P^{(1)}_{e,n}(\mathcal{C})} + \frac{1}{|\mathcal{C}|},\label{eq:P-rank-C}
\end{align}
by using the union bound and since by symmetry $P^{(1)}_{e,n}(\mathcal{C}) = P^{(2)}_{e,n}(\mathcal{C}) \eqqcolon P_{e,n}(\mathcal{C})$. Without loss of generality, we therefore focus on the individual~(and symmetric) probability of \emph{decoding error}\footnote{
	Note that this probability differs slightly from the classical decoding error $\mathbb{P} {\big\{ M_j \ne \hat{M}_j \given \mathcal{C} \big\}}$, since we condition on $M_1\ne M_2$.
}
$P_{e,n}(\mathcal{C})$.
We also have the lower bound
\begin{equation} 
	\mathbb{P} \left\{ Z_1 = Z_2 \given \mathcal{C} \right\}
	\ge  \mathbb{P} \left\{ M_1 = M_2 \given \mathcal{C} \right\}
	=  \frac{1}{2 |\mathcal{C}|},
\end{equation}
since $M_1 = M_2$ implies $(\Xv_{1},\Yv_{1}) = (\Xv_{2},\Yv_{2})$ and if this occurs, the optimal strategy is to select identifiers uniformly at random, resulting in a synchronization error with probability $1/2$. 

%% file: 3_random_codes.tex
\section{Random Codes and Error Analysis}
\label{sec:random_codes}

In Section~\ref{sec:bandit_channel_model}, the decoding error probability $P_{e,n}(\mathcal{C})$ was defined for a fixed codebook $\mathcal{C}$; see~\eqref{eq:P-rank-C}.
In this section, we investigate which classes of codebooks minimize this error probability.
A standard approach to analyzing decoding error probabilities is to average them over an ensemble of random codes~\cite{gallager1968}. This technique establishes the existence of at least one code whose error probability is no larger than the ensemble average. Moreover, the choice of the random coding distribution provides insights on the structure of codebooks that achieve low decoding error.

We consider two ensembles of \emph{block random codes}, both constructed as concatenations of $N_B$ subblocks. Specifically, we fix a partition
\[
\Bv \coloneqq (B_1, \dots, B_{N_B})
\]
of the block $[n]$, where $B_b \subseteq [n]$ for each $b \in [N_B]$,
$\bigsqcup_{b \in [N_B]} B_b = [n]$, and the $b$-th subblock has length $n_b \coloneqq |B_b|$, with $\sum_{b \in [N_B]} n_b = n$. Codewords are generated independently across subblocks, and each subblock $B_b$ is associated with a distribution $Q_b \in \mathcal{P}([K])$.

A block random coding ensemble is therefore fully specified by the partition $\Bv$ and the collection of distributions
\[
\Qv \coloneqq (Q_1, \dots, Q_{N_B}).
\]
This construction is analogous to \emph{coded time-sharing}~(see, e.g.,~\cite{el_gamal_network_2012}) except that here the time-sharing variable is not generated randomly.

We consider the following two variants.
\begin{itemize}
	\item In \emph{i.i.d. block random coding}, the symbols within each subblock $B_b$ are generated independently and identically according to $Q_b$, and the subblocks are independent across $b$.  The probability that message $m$ is encoded as a sequence $\xv \in [K]^n$ is given by
\begin{equation}
Q^{\Biid}_{\Xv}(\xv)
\coloneqq
\prod_{b \in [N_B]} \prod_{i \in B_b} Q_b(x_i).
\label{eq:iid_block_code_distribution}
\end{equation}

	\item In \emph{constant-composition block random coding}, the subsequence corresponding to each subblock $B_b$ is drawn uniformly at random from the type class
$\mathcal{T}_{n_b}(Q_b)$, and different subblocks are chosen independently.
Thus, the probability that message $m$ is encoded as $\xv \in [K]^n$ is
\begin{equation}
Q^{\Bcc}_{\Xv}(\xv)
\coloneqq
\prod_{b \in [N_B]}
\frac{1}{\lvert \mathcal{T}_{n_b}(Q_b) \rvert}
\,
\mathbbm{1}\!\left\{ \xv_{B_b} \in \mathcal{T}_{n_b}(Q_b) \right\}.
\label{eq:cc_block_code_distribution}
\end{equation}
By definition, a sequence $\xv_{B_b}$ belongs to $\mathcal{T}_{n_b}(Q_b)$ if and only if
\[
\sum_{i \in B_b} \mathbbm{1}\{x_i = x\}
= n_b Q_b(x),
\qquad \forall\, x \in [K].
\]
\end{itemize}

In particular, we consider two specific choices of block partitions and associated distributions. In \emph{single-block uniform (U) random coding}, we take $N_B=1$ with a single block $B_1 = [n]$, and assign the uniform distribution $Q_1(x) = 1/K$, for each $x\in[K]$. In \emph{combinatorial-pair~(CP) block uniform random coding}, we take $N_B = \binom{K}{2}$ subblocks, each of length $\lfloor n/N_B \rfloor \le n_b \le \lfloor n/N_B \rfloor + 1$. Further, we assign to each subblock a distinct (unordered) pair of symbols $\Acal_b \coloneqq \{ x_{b,1}, x_{b,2} \} \subseteq [K]$, and the corresponding distribution $Q_b(x) = \frac{1}{2} \1\left\{ x \in \Acal_{b} \right\}$.

 We denote by $\bar{P}_{e,n} \coloneqq \bar{P}_{e,n}(\Bv,\Qv)$ the \emph{average probability of error} over the ensemble of block random codes defined by the partition~$\Bv$ and distributions~$\Qv$.%

\begin{theorem}[General probability of error]
		For i.i.d. block random coding with $N_B$ blocks, defined by partition $\Bv$ and distributions $\Qv$, the average probability of error under ML decoding satisfies
	\begin{equation*}
		\bar{P}_{e,\ML}(\Bv,\Qv)(n) \le \min_{\rho\in[0,1]} (M-2)^\rho \prod_{b=1}^{N_B} \Big( \sum_{x=1}^K F_\rho\big( Q_b(x),\mu_x \big) \Big)^{n_b} \, ,
		\label{eq:proba_of_error}
	\end{equation*}
	where 
    \begin{equation}
    F_{\rho}(q, \mu) \coloneqq q(1-q)^{1+\rho} \rw + q \left[ q + (1-q)(1-\rw)^{\frac{1}{1+\rho}} \right]^{1+\rho}.
    \end{equation}
    \label{thm:probability of error}
\end{theorem}
\vspace{-2em}
\begin{IEEEproof}
    The proof follows standard random-coding arguments \cite{gallager1968} adapted to the bandit channel, see \cite[Thm.~2]{zhang2025}. A key difference in the present setting is that the two transmitted messages are distinct and that the random code is generated independently across subblocks.
\end{IEEEproof}

\begin{theorem}[Achievable rate]
Consider i.i.d.~block random coding with $N_B$ subblocks, specified by a partition $\Bv$ and a collection of input distributions $\Qv$, with subblock lengths $n_b$ for each $b \in [N_B]$.
A rate $R$ is achievable, i.e., the decoding error probability in~\eqref{eq:proba_of_error} vanishes as $n \to \infty$, if
\begin{equation}
R < R^\star(\Bv,\Qv)
\coloneqq
I(Y; X_2 \mid X_1, U),
\end{equation}
where the joint distribution of $(U, X_1, X_2, Y)$ is given by
$P_U P_{X_1 X_2 Y \mid U}$,
with conditional distribution
\[
p(x_1,x_2,y \mid u)
=
Q_u(x_1)\, Q_u(x_2)\, P(y \mid x_1,x_2),
\]
and where $p_U(b) = n_b/n$ for each $b \in [N_B]$.

\end{theorem}

\begin{IEEEproof}
    Similarly to \cite[Thm.~4]{zhang2025}, and considering the expression of the average probability of error in Theorem~\ref{thm:probability of error},
    we can derive the achievable rate by considering the limit when $\rho$ goes to $0$ of $-\frac{1}{\rho}\sum_n^{N_b} l_b  \log \left( \sum_{x=1}^K F_\rho\left(Q_b(x), \mu_x\right)\right)$, which gives the mutual information averaged over each block. 
\end{IEEEproof}

\begin{cor}[Probability of error for single-block and CP-block codes]
	Define $\kappa_{\mu_x} \coloneqq 1-\sqrt{1-\mu_x}$, for all $x \in [K]$.
	
	For the single-block uniform random code ensemble, the average probability of error under ML decoding satisfies
	\begin{equation*}
		\bar{P}_{e,\ML}^{\text{U}}(n) \leq (M-2)e^{-nE^{\text{U}}(\mu)+1},
	\end{equation*}
	where the error exponent $E^{\text{U}}(\mu)$ is given by
	\begin{equation}
		E^{\text{U}}(\mu) \coloneqq -\log \left(1 - \frac{2}{K^2}\left(1-\frac{1}{K}\right) \sum_{x\in[K]} \kappa_{\mu_x}\right).
	\end{equation}
	
	For the CP-block uniform random code ensemble, the average probability of error under ML decoding satisfies
	\begin{equation*}
		\bar{P}_{e,\ML}^{\text{CP}}(n) \leq (M-2)e^{-nE^{\text{CP}}(\mu)+1}
	\end{equation*}
	where the error exponent $E^{\text{CP}}(\mu)$ is
	\begin{equation}
		E^{\text{CP}}(\mu) \coloneqq -\log \left(1 - \frac{1}{2K}\sum_{x \in [K]} \kappa_{\mu_x} \right).
	\end{equation}
\end{cor}

\begin{IEEEproof}
	We take the expression in Theorem~\ref{thm:probability of error}, set $\rho = 1$, and consider the particular block structure and distribution of each code.
\end{IEEEproof}

There is a multiplicative gap of at least $K/2$ between the two error exponents, which can be explained by the number of collisions induced by each ensemble. Under CP random coding, the expected number of collisions during synchronization is $n/2$, whereas under single-block uniform random coding it is only $n/K$, which becomes negligible for large $K$. Since collisions are necessary for communication under the bandit channel model, the CP construction induces a controlled, non-vanishing number of collisions by restricting each subblock to a pair of arms, thereby enabling reliable synchronization.

%% file: 4_universal_decoder.tex
\section{Universal Decoder for the Bandit Channel}
\label{sec:universal_decoder}

In this section, we propose decoding strategies for the bandit channel that, unlike the ML decoding rule, do not rely on the knowledge of the channel parameters~$\mu$. We focus on constant-composition CP-block random coding, based on our error-exponent analysis in Section~\ref{sec:random_codes}. 
For a sequence $\xv\in\Xcal^n$, we denote $\xv_b \coloneqq \left( x_i \colon i \in B_b \right)$ the symbols in the $b$-th block, so that $\xv \cong \left( \xv_1, \dots, \xv_{N_B} \right)$ (analogously for $\xv_1$, $\xv_2$, $\yv$ etc.).

In each block, one can compute the empirical conditional distribution
\begin{align}
	&\hat{P}_{\xv_{2,b} \given \xv_{1,b}, \yv_{b}}(x_2 \given x_1, y) \nonumber\\
	&\hspace{3em} = \frac{\sum_{i\in B_b} \1\{ (x_{i,1}, x_{i,2}, y_i)=(x_1,x_2,y) \} }{ \max(1,\sum_{i\in B_b} \1\{ (x_{i,1}, y_i)=(x_1,y) \} ) },
\end{align}
which equals $0$ if the denominator is~$0$. We can then consider the decoding rule~\eqref{eq:decoding-rule} with the metric
\begin{equation} \label{eq:universal-metric}
	u_m\left( \xv_1, \xv_2, \yv \right) \coloneqq 2^{- \sum_{b\in[N_B]} n_b H \left( \hat{P}_{\xv_{2,b} \given \xv_{1,b}, \yv_{b}} \right)}.
\end{equation}
Note that this function does not depend on the channel parameters~$\mu$.
This decoding rule is similar to the minimum conditional entropy rule of~\cite{ziv1985}, except that empirical entropies are computed block-wise, and then weighted with the lengths of each block.
The next result states that the exponent of the average probability of error with this decoding rule is essentially the same as with the ML rule.

\begin{theorem} \label{thm:universality}
	Let $\smash{\bar{P}^{\text{CP}}_{e,m}(n)}$ and $\smash{\bar{P}^{\text{CP}}_{e,\ML}(n)}$ denote the average probability of error for constant-composition CP-block random coding under decoding with metric m and with the ML metric, respectively. Then,
		\begin{equation}
			\bar{P}^{\text{CP}}_{e,m}(n) \le \bar{P}^{\text{CP}}_{e,\ML}(n)\, 2^{n\delta(n,K)},
		\end{equation}
		where
		\begin{equation}
			\delta(n,K) \coloneqq \frac{2K^3(K-1)}{n} \log \left( \frac{2n}{K(K-1)}+2 \right),
		\end{equation}
		where $\delta(n,K) \to 0$ as $n\to\infty$ for any fixed $K$.
\end{theorem}
\begin{IEEEproof}
	The proof is based on the method of types~\cite[Ch.~2]{csiszar2011}, and follows the same lines of \cite{ziv1985,feder1998} (see also~\cite{miyamoto2024}), with the difference that empirical entropies are computed block-wise.
\end{IEEEproof}

%% file: 5_application_to_bandit.tex
\section{Regret Analysis}
\label{sec:application_to_bandit}

Using the results from the previous sections, we can now analyze the regret of the initialization phase of a distributed multi-player bandit algorithm that uses our proposed synchronization scheme.

\begin{theorem}[Synchronization regret as a function of synchronization length]
    There exists a code of length $n$ so that for any $\mu$ the regret of the synchronization phase using the universal decoder is bounded by
    \begin{equation*}
    \begin{split}
    \mathcal{R}_{\text{init}} \le\;& n(\mu_{(1)}+\mu_{(2)}) \\
    &\hspace{-1em}+ T(\mu_{(1)}+\mu_{(2)})\left(M2^{-n\big(E^{\mathrm{CP}}(\mu)-\delta'(n,K)\big)}+M^{-1}\right)\,,
    \end{split}
    \end{equation*}
    where $\delta'(n,K) \coloneqq \hq{\delta}(n,K) + \frac{1}{n}$; thus,  $\lim_{n \to + \infty}\delta'(n,K) = 0$.
    \label{thm:regret_bound}
\end{theorem}

\begin{cor}[Asymptotic regret of synchronization]
Setting $M = T^{1+ \epsilon}$ and $n = 2(1+\epsilon)\frac{1}{E^{\text{cp}}(\mu)} \log(T)$, we have that the asymptotic regret of the synchronization phase, using the universal decoder with block uniform random code, satisfies
    \begin{equation}
       \lim_{T \to + \infty} \frac{\mathcal{R}_{\text{init}}(\mu,T)}{\log(T)} \leq 2(1+\epsilon) \frac{(\mu_{(1)} + \mu_{(2)})}{E^{\text{CP}}(\mu)} \leq 8(1+\epsilon)K   \, .
    \end{equation}
    \label{cor:regret_bound}
\end{cor}

The ratio between the best reward and the error exponent ${(\mu_{(1)} + \mu_{(2)})}/{E^{\text{CP}}(\mu)} \leq 4K$ is upper bounded. In fact, if we define $\tilde{\mu} \coloneqq (\mu_{(1)}, 0,\ldots,0)$ we have that $E^{\text{CP}}(\mu) \geq E^{\text{CP}}(\tilde{\mu})$ and $(\mu_{(1)} + \mu_{(2)})/E^{\text{CP}}(\tilde{\mu}) \leq 4K$. Therefore, knowing $\mu_{(1)}$ alone allows us to set the initialization length to have a regret scaling in $8(1+\epsilon)K \log(T)$. 

The published conference version of this paper claimed that this scaling is minimax optimal in terms of $K$ and $\log(T)$, based on Theorem~\ref{thm:minimaxoptimality} below. \textcolor{red}{This claim is withdrawn:} the theorem is incorrect, as explained in the footnote to the abstract. Its statement and proof sketch are kept only for reference, and they should not be used.

\begin{theorem}[\textcolor{red}{INCORRECT} -- Minimax regret of synchronization]
    For any synchronization scheme using the empirical distribution of $\mu$ as prior information we have 
    \label{thm:minimaxoptimality}
\begin{align}	
	 \lim\inf_{T \to \infty} \max_{\mu }\frac{\mathcal{R}_{\text{init}}(\mu,T)  }{\log(T)} \ge \frac{K}{8 \log(2) } \left(1 - \frac{2}{K}\right).
\end{align}
\end{theorem}
\begin{IEEEproof}
    We fix $m \in (0,1)$ and consider the prior on $\mu$, with $\mu$ drawn according to a distribution such that, $k^\star$ is chosen uniformly at random in $[K]$ and $\mu$ is chosen as a function of $k^\star$ such that $\mu_k = m \mathbbm{1}\left\{ k = k^\star\right\}$. We then compute, for a fixed $\mu \in \Theta$, the probability that both players only see zeros during the initialization phase of length $n$. Averaging the regret caused by this event over the prior, we get the result.  
\end{IEEEproof}

%% file: 6_numerical_results.tex
\section{Numerical Results}
\label{sec:numerical_results}

In this section, we present numerical results illustrating the performance of the proposed synchronization scheme for distributed multi-armed bandits using the bandit channel. Throughout this section, we fix the codebook size to $M = 10^4$ and use between $5 \times 10^4$ and $10^5$ Monte Carlo trials to estimate the probability of error.

\begin{figure}
  \begin{center}
    \includegraphics[width=0.40\textwidth]{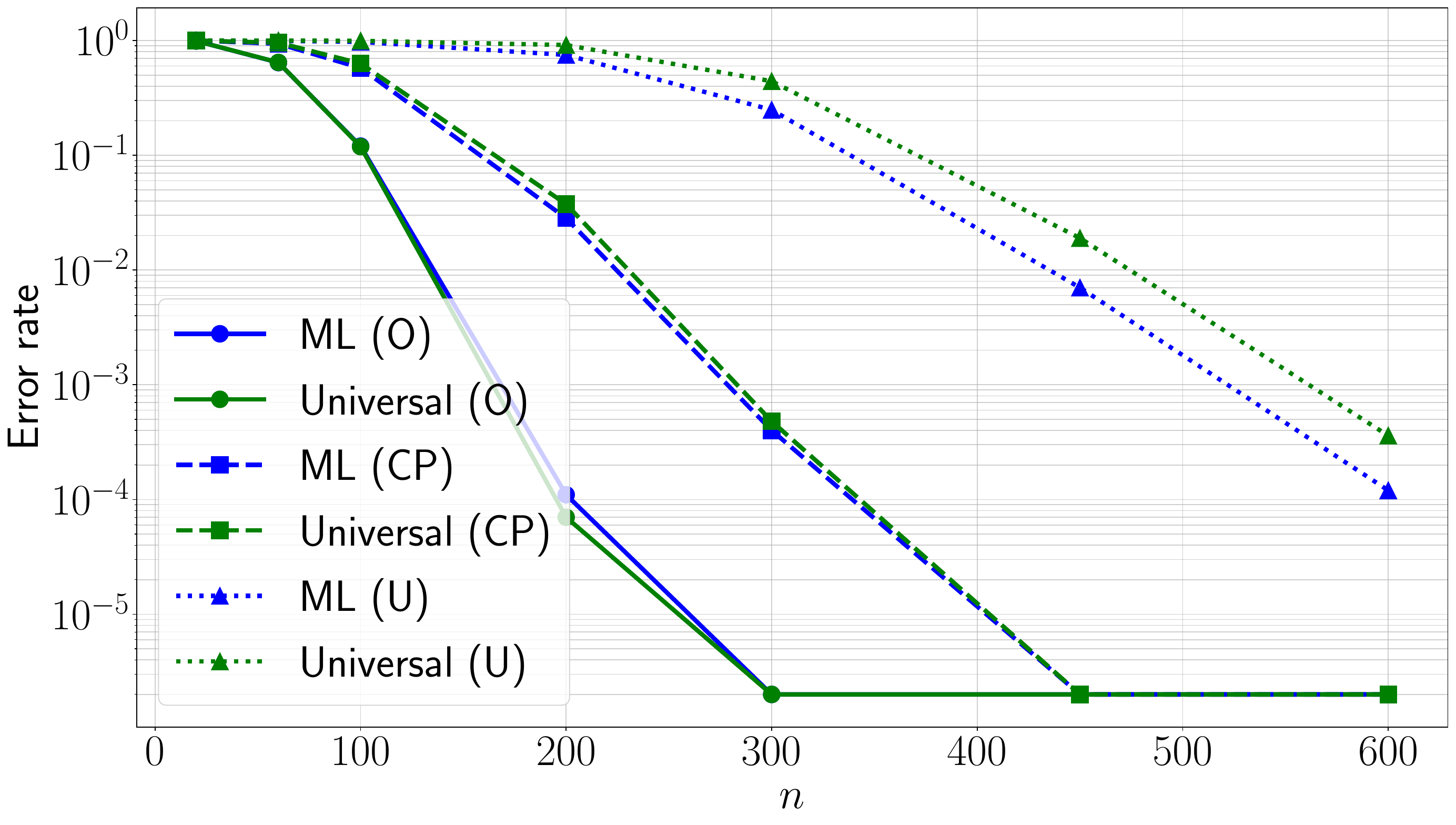}
  \end{center}
  \caption{Communication error vs. blocklength $n$.}
  \label{fig:comunication_error}
\end{figure}

We first compare the performance of the proposed coding and decoding schemes, as shown in Fig.~\ref{fig:comunication_error}. We consider a bandit channel with $K = 6$ arms and parameters
$\rw = [0.3,\, 0.3,\, 0.3,\, 0.1,\, 0.1,\, 0]$.
Three codebook distributions are evaluated: the optimal distribution (O) from~\cite{zhang2025}, the combinatorial-pair (CP) distribution, and the uniform (U) distribution. For each blocklength $n$, a single codebook is generated according to each distribution.
As predicted by the theory, the universal decoder performs close to the ML decoder for all codebook distributions. The combinatorial-pair codebook significantly outperforms the uniform codebook. The optimal codebook achieves the best performance overall, but requires exact knowledge of the channel parameters.

\begin{figure}
  \begin{center}
    \includegraphics[width=0.40\textwidth]{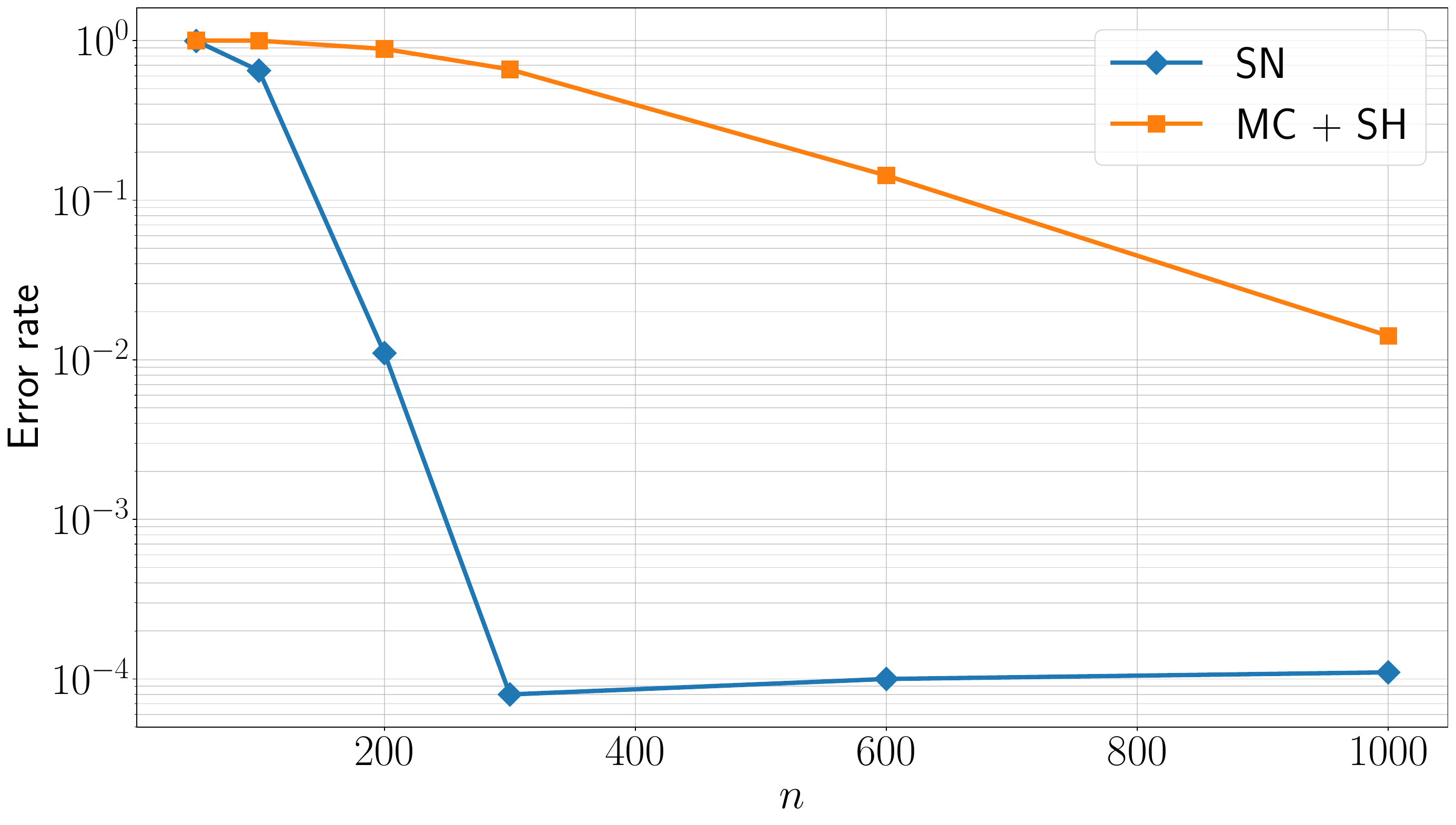}
  \end{center}
  \caption{Synchronization error as a function of $n$.}
  \label{fig:sync_error_n}
\end{figure}

We next compare our synchronization scheme, referred to as \emph{SendNumber} (SN), with the Musical Chairs + Sequential Hopping (MC+SH) scheme from~\cite{boursier_sic-mmab_2019}, as shown in Fig.~\ref{fig:sync_error_n}. We consider a bandit channel with $K = 6$ arms and parameters
$\rw = [0.35,\, 0.3,\, 0.25,\, 0.2,\, 0.15,\, 0.1]$,
for various blocklengths $n$.
For MC+SH, the number of repetitions is set to $T_C = n/(3K)$, following~\cite{boursier_sic-mmab_2019}.
As $n$ increases, the error probability of SN decays much faster than that of MC+SH and rapidly approaches the baseline level $1/M$.

\begin{figure}
  \begin{center}
    \includegraphics[width=0.40\textwidth]{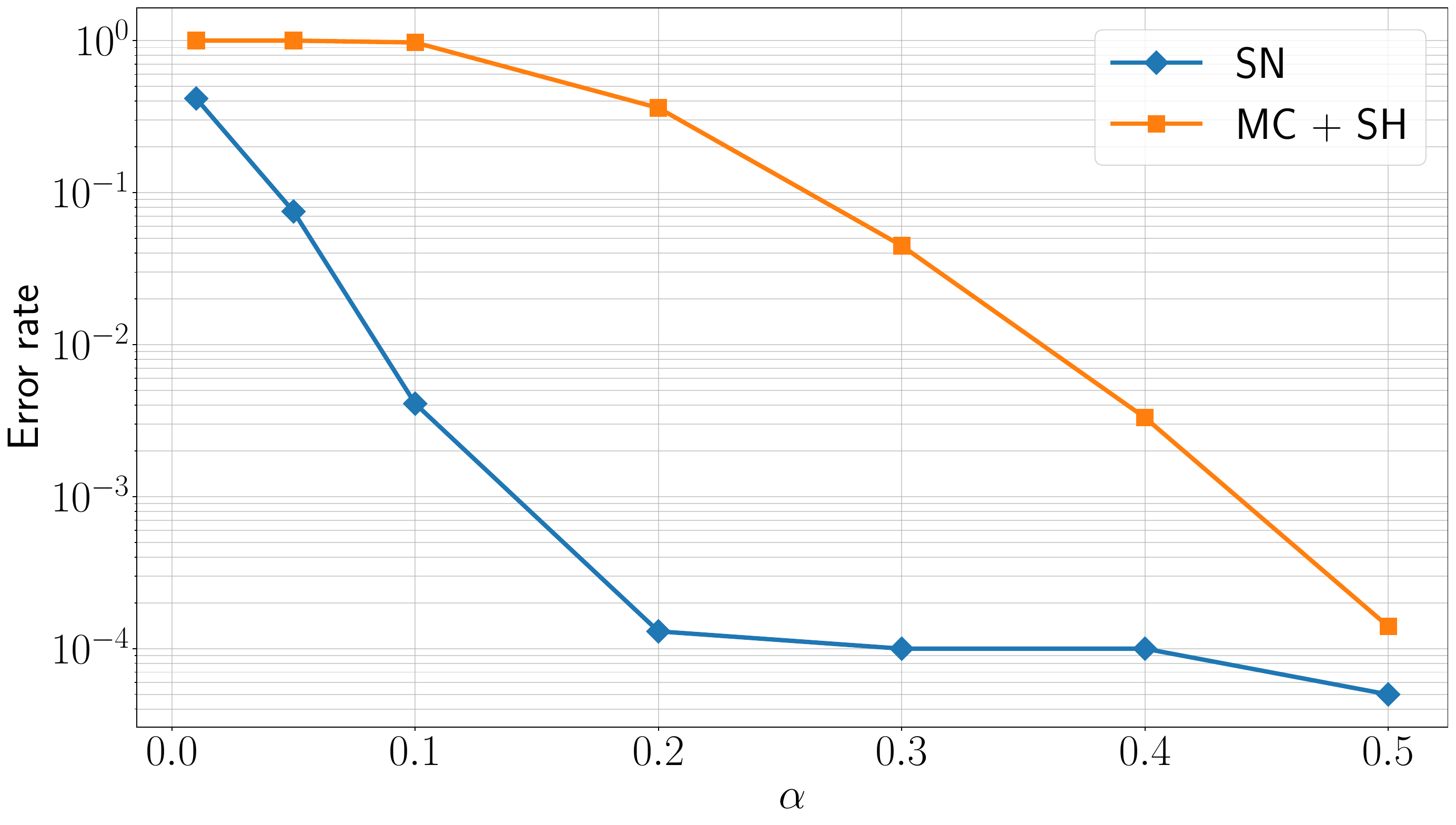}
  \end{center}
  \caption{Synchronization error as a function of $\alpha$.}
  \label{fig:sync_error_alpha}
\end{figure}

Finally, we compare the performance of the schemes as one channel parameter tends to zero, as shown in Fig.~\ref{fig:sync_error_alpha}. We consider a bandit channel with $K = 6$ arms and parameters
$\rw = [0.5,\, \alpha,\, \alpha,\, \alpha,\, \alpha,\, \alpha]$,
where $\alpha$ varies from $0$ to $0.5$. The initialization length is fixed to $n = 300$.
As $\alpha$ decreases, the error probability of MC+SH converges to one and increases significantly faster than that of SN which is strictly less than 1 even when $\alpha = 0$. As $\alpha$ increases toward $0.5$, the error probability of SN rapidly approaches the lower bound of $1 / M$.

\section{Conclusion}
We proposed a synchronization scheme for the distributed multi-armed bandit problem with two players. The scheme leverages the bandit channel model, a combinatorial-pair codebook, and universal decoding techniques. We derived an upper bound on the regret of the resulting synchronization scheme that scales asymptotically as $K \log(T)$ in the number of arms $K$ and the time horizon $T$. Numerical results further illustrate the effectiveness of the proposed approach compared to the state of the art.